\documentstyle[prd,aps,preprint,tighten,epsfig]{revtex}
\begin{document}
\preprint{                                           BARI-TH/364-99}
\draft
\title{ Testing solar neutrino MSW oscillations at low $\delta m^2$ \\
	through time variations of event rates in GNO and BOREXINO}
\author{
G.L.\ Fogli~$^a$, E.\ Lisi~$^a$, D.\ Montanino~$^b$, and A.\ Palazzo~$^a$}
\address{     $^a$~Dipartimento di Fisica and Sezione INFN di Bari,\\
                   Via Amendola 173, I-70126 Bari, Italy}
\address{$^b$~Dipartimento di Scienza dei Materiali dell'Universit\`a di Lecce,\\
             Via Arnesano, Collegio Fiorini, I-73100 Lecce, Italy}
\maketitle
\begin{abstract}
The Mikheyev-Smirnov-Wolfenstein (MSW) explanation of the solar neutrino
problem is currently compatible with three distinct regions of the two-neutrino
oscillation parameter space $(\delta m^2,\sin^2 2\theta)$.  We focus on the
region with the lowest value of $\delta m^2$ ($\sim 10^{-7}$ eV$^2$), which
implies significant Earth regeneration effects for low-energy solar neutrinos.
We point out that such effects are not only observable as {\em day-night\/}
variations of neutrino event rates in the real-time BOREXINO experiment, but
also as {\em seasonal\/} variations in the radiochemical Gallium Neutrino
Observatory  (GNO) at Gran Sasso. We present detailed calculations of the
difference between winter and summer rates in GNO (six months averages) in
excess of the trivial seasonal variation due to the Earth orbital eccentricity.
We show that, within  the low-$\delta m^2$ MSW solution, the net winter-summer
GNO rate difference amounts to 4--6  SNU, with a dominant contribution from
{\em pp\/} neutrinos. We also give analytical expressions for the winter and
summer solar exposure functions at the Gran Sasso site.
\end{abstract}

\section{Introduction}

The available solar neutrino data from the Homestake (Cl)  \cite{Cl98}, GALLEX
(Ga) \cite{Ha99},  SAGE (Ga) \cite{Ab99},  Kamiokande (K) \cite{Fu96},  and
Super-Kamiokande (SK) \cite{To99,Su99} experiments indicate consistently a
significant $\nu_e$ flux deficit \cite{NuAs} with respect to the predictions of
the Standard Solar Model (SSM) \cite{BP98}.  The Mikheyev-Smirnov-Wolfenstein
(MSW) mechanism of neutrino oscillations in matter \cite{MSWm} represents a
possible explanation of this deficit.

Recent global analyses of the data within the MSW mechanism  for active $2\nu$
oscillations \cite{Su99,700d,Last} currently select three separate regions in
the  parameter space $(\delta m^2,\sin^22\theta/\cos 2\theta)$,  as shown in
Fig.~1.%
\footnote{The results in Fig.~1 are derived by our analysis of solar $\nu$
	data, updated as of summer~'99 conferences. The details are not
	relevant for the present work and are discussed elsewhere 
	\protect\cite{Last}.}
The three allowed regions are usually indicated \cite{Ba97} as ``small mixing
angle'' (SMA), ``large mixing  angle'' (LMA), and ``low $\delta m^2$'' (LOW)
solutions.

The precise shape of the allowed regions in Fig.~1 is subject to changes due to
fluctuations in the data, improvements in the SSM fluxes, and details of the
statistical analysis. In particular, the LOW solution, which tends to
overestimate the Cl rate and to underestimate the Ga rate, is a ``borderline''
case: it may be allowed (see, e.g., \cite{Su99,700d,Last,Ba97,Ha93,3MSW,Ba94}) 
or excluded (see, e.g., \cite{Ca95,100d,300d,500d})  at 99\% C.L., as  a
consequence of relatively small variations in the data or in the theoretical
predictions and their uncertainties. Notice that the recent {\em spectral\/}
data from SK \cite{To99,Su99} do not show  clear deviations from standard
expectations, and are compatible with the LOW solution, which predicts small or
null distortions of the SK time-energy distributions of events (see, e.g.,
Figs.~5--7 of Ref.~\cite{Li97} for $\delta m^2\sim 10^{-7}$).

Clearly, the status of the LOW solution (as well as of the SMA and LMA
solutions) can be firmly assessed only through more precise or new experimental
data. Since the MSW physics scales as $\delta m^2/E_\nu$, the low-$\delta m^2$
region is more properly tested through observations of the low-energy part of
the solar neutrino spectrum. In particular, within the LOW solution the  Earth
regeneration effects \cite{Sm85} are rather large for Be and {\em pp\/} solar
$\nu_e$'s (see, e.g., \cite{Ba94} and Fig.~9 of \cite{500d}).  Such effects can
be revealed as time variations of event rates in two, new-generation solar
neutrino experiments at Gran Sasso: BOREXINO \cite{BORE}  (based on $\nu$-$e$
scattering) and the Gallium Neutrino Observatory (GNO) \cite{GNOE}  (based on
$\nu_e$-Ga absorption). However, while the real-time experiment BOREXINO can
observe {\em day-night\/} rate variations, the radiochemical  GNO experiment
can  detect only long-term, {\em seasonal\/} variations related  to the longer
neutrino pathlength in the Earth during winter.%
\footnote{ See \protect\cite{Sm99,Seas,Ba99}  for recent discussions and
	earlier references about MSW-induced seasonal effects. Extensive
	references to day-night effects can be found in \protect\cite{Li97}.}

In this work, we first revisit (Sec.~II) the calculation of the night-day
asymmetry expected in BOREXINO. Then we present (Sec.~III) a detailed
calculation of the  winter-summer rate difference $R_W-R_S$ (six months
averages) in GNO, with eccentricity effects removed. We show that, within the
LOW solution, one expects $R_W-R_S\simeq 4$--6 SNU, with a dominant
contribution from  {\em pp\/} neutrinos. We conclude our work in Sec.~IV, and
give in the Appendix analytical formulae useful for winter and summer averages
at the Gran Sasso site.

\section{Day-night variations in BOREXINO}

Figure~2 shows our calculation of the event rates (normalized to  SSM
unoscillated values) in BOREXINO during daytime (D)  and nighttime (N),
averaged over the year. The scales are the same as in Fig.~1. In particular, we
adopt $\sin^2 2\theta/\cos 2\theta$ as abscissa, rather than the usual variable
$\sin^2 2\theta$, in order to expand the plot in the region at large mixing,
interesting for the LOW solution. The calculation includes updated neutrino
fluxes, spectra,  and production regions in the Sun \cite{BP98},   radiatively
corrected differential cross sections \cite{Ka95} for $\nu_e$ and for the
oscillating partners $\nu_{\mu,\tau}$, and BOREXINO detector technical
specifications as in \cite{Fa99}. The $\nu_e$ survival probability  is
calculated taking into account an accurate Earth density profile \cite{PREM},
and  is then averaged over nighttime as described in \cite{Li97}.  Each panel
of Fig.~2 is sampled through a $100 \times  100$ grid.

The comparison of the  yearly-averaged nighttime and daytime rates in Fig.~2 
shows that day-night variations larger than 1\% are expected in BOREXINO over
three decades in $\delta m^2$ ($\sim 10^{-8}$--$10^{-5}$ eV$^2$),  as a
consequence of (mainly Be) neutrino regeneration in the Earth. The maximum
night-day asymmetry can be as large as $\sim 40\%$  at $(\delta m^2/{\rm
eV}^2,\sin^2 2\theta)\simeq (3\times 10^{-7},0.2)$  (similar results have been
obtained in \cite{Ba97,Mu99}). However, within the LOW solution of Fig.~1, the
values expected  for the asymmetry are smaller ($N-D/N+D\simeq 10$--20\%).
Although a 10\% asymmetry seems to be sufficiently large for a clear detection
in BOREXINO \cite{Mu99}, this fact indicates that  the Be neutrino energy is
somewhat too high to test the LOW region with full-strength Earth regeneration
effects. Therefore, the logical next step is to investigate the Earth
regeneration effect in experiments sensitive to the lowest-energy ({\em pp})
neutrinos, namely, gallium absorption experiments.

\section{Seasonal variations in GNO}

When the first solar neutrino results from GALLEX and SAGE became available,
the authors of  \cite{Ba94} pointed out that large {\em day-night\/} variations
should be expected in such experiments within the LOW solution (which they
called ``region C''). However,  such daily variations of the Ga event rate
\cite{GaND}  are practically unobservable, being averaged out in each
experimental run.%
\footnote{Future  real-time observations of day-night effects for {\em pp\/}
	neutrinos might be possible through a recently proposed technique
	\protect\cite{Ra99} based on  $\nu_e$ absorption on  Yb and Gd targets,
	as in the Low Energy Neutrino Spectroscopy (LENS) project
	\protect\cite{LENS}. Other projects (HERON, HELLAZ) focus on  $\nu$-$e$
	scattering in Helium  \protect\cite{Heli}. Such proposals are not
	investigated in this work.}
Similarly, day-night effects for B and Be neutrinos are  effectively smeared
out in the Homestake (Cl) experiment \cite{Cl98},  since early proposals of
separate day-night extractions \cite{ClND} have never been realized.
Nevertheless, it was noted in several 1986-87 papers that ``... a day-night
difference, if it is sufficiently large, should reveal itself also in a
summer-winter difference'' \cite{Da97} (see also \cite{Sm85,Hi87}).  However, 
given the limited statistics of the  first-generation Ga and Cl experiment,
the   possibility of observing MSW-induced seasonal variations in radiochemical
experiments was not (to our knowledge) further investigated.

Here we fill this gap by showing that such possibility  might become realistic
in the second-generation, high-statistics GNO experiment at Gran Sasso, which
should significantly improve the already excellent performance of its parent
experiment GALLEX.  In fact, we estimate  that MSW-induced {\em seasonal\/}
variations  in the range $\sim 4-6$ SNU are expected in a Ga experiment (within
the LOW solution). Such variations,  which are below the GALLEX and SAGE 
detection sensitivities \cite{Ca95}, might well be observed at GNO,  where the
prospective statistical error after one solar cycle amounts to $\lesssim 2$ SNU
\cite{GNOE}.  Indeed, the observation of the purely geometrical ($1/L^2$)
seasonal variations due to the Earth orbit eccentricity represents one of the
possible goals in GNO \cite{GNOE}.

Being interested only in the nontrivial seasonal variations, we remove
eccentricity effects and focus on the net effects due to $\nu_e$ regeneration
in the Earth (see the Appendix for further details). We conventionally divide
the year in two six-months periods (indicated as ``Winter'' and ``Summer'') 
centered at the solstice days and separated by the equinox days:
\bigskip
\bigskip
\begin{eqnarray}
{\rm Winter} &\equiv& {\rm winter\ solstice\ (21\ dec.)}\;\pm 3{\rm\ months}
\nonumber\\
&\simeq & [{\rm 23\ september,~21\ march}]\ ,\\
\nonumber\\
{\rm Summer} &\equiv& {\rm summer\ solstice\ (21\ jun.)}\;\pm 3{\rm\ months}
\nonumber\\
&\simeq & [{\rm 22\ march,~22\ september}]\ .\\
\nonumber
\end{eqnarray}
The neutrino oscillation probabilities and event rates are then averaged over
such periods through a ``weight function'' approach similar to that of
Ref.~\cite{Li97}, as discussed in the Appendix.

Figure~3 shows the results of our calculation of the gallium absorption rates
(in SNU) averaged over winter $(R_W)$, summer $(R_S)$, and year $(R_Y)$,
together with the main new result of this work, the winter-summer difference
$R_W-R_S$. Since eccentricity effects are removed, the difference $R_W-R_S$
approaches zero in the limits of no oscillations ($\delta m^2\to 0$ or
$\theta\to 0$) and of energy-averaged oscillations  (large $\delta m^2$), up to
small round-off errors (see below). The comparison of the two upper panels 
($R_W$ and $R_S$) shows that, in the region of the LOW solution ($\delta
m^2\sim 10^{-7}$), large Earth regeneration effects distort the iso-SNU Gallium
curves both in winter and in summer; indeed, the corresponding distortion in
the $R_Y$ panel is partly responsible for the appearance of the LOW solution in
fits to the data. Notice that, in the region of the LOW solution, the
yearly-averaged theoretical rate is $R_Y\sim 60$ SNU, which is consistent with
the SAGE results ($\sim 67\pm 8$ SNU \cite{Ab99}), but somewhat underestimates
the GALLEX rate ($\sim 77.5 \pm 8$ SNU \cite{Ha99}). Therefore, the first check
of the LOW solution in GNO will come from a more precise measurement of the
total rate averaged over the year.

In the fourth panel of Fig.~3, the difference $R_W-R_S$ shows a rather
complicated structure as a  function of the mass-mixing parameters, with
several peaks and valleys.  The values of $R_W-R_S$ are  almost everywhere
positive, since Earth matter effects are typically less pronounced in summer
than in winter, when the neutrino trajectories probe the inner part of the
Earth and can cross its core,%
\footnote{Recent analytical studies of  neutrino oscillations along the Earth
	mantle-core density profile are presented in  \protect\cite{CoMa}.}
as shown in the Appendix. The maximum value of $R_W-R_S$ is $8.3$ SNU. However,
small negative values ($-0.7$ SNU at minimum) can be reached around  $(\delta
m^2/{\rm eV}^2,\sin^2 2\theta/\cos 2\theta)\simeq(10^{-7},0.7)$, inside the
``boomerang-shaped'' region labeled by zero. Smaller negative values (tenths of
a SNU) can also be reached around  $(\delta m^2/{\rm eV}^2,\sin^2 2\theta/\cos
2\theta) \simeq(1.5\times10^{-7},0.003)$.%
\footnote{Similarly, night-day rate differences can take slightly negative
	values in some regions of the parameter space
	\protect\cite{Ba97,Pe98}.}
The outermost curve, labeled $\sim 0$ SNU, crosses a region at very low
gradient, and is thus numerically unstable: very small changes (percents of a
SNU) can shift it significantly. All the other isolines are stable, being
located in high-gradient regions. Since we estimate  a numerical accuracy of
about $\lesssim 0.03$ SNU in our total rates, the $\sim 0$ SNU curve simply
indicates a zone with negligibly small  winter-summer difference. This zone
covers the SMA and LMA regions of Fig.~1. Negative or zero values for the
winter-summer difference are instead  irrelevant for the LOW solution region,
where the difference $R_W-R_S$ is positive and greater than about 4 SNU, with a
local maximum of 6.3 SNU. Therefore, if the LOW solution is the explanation of
the solar neutrino problem, one should observe in GNO a net winter-summer
difference (in excess of the trivial eccentricity difference) of about  4 to 6
SNU. No significant difference should be observed in the case of the SMA or LMA
solutions.

In order to understand better the structure of the $R_W-R_S$ isolines, we show
in Fig.~4 the separate contributions to this difference coming from the {\em
pp}, Be, {\em pep}+CNO, and B solar neutrino fluxes.%
\footnote{In Fig.~4,  the isolines labeled ``$\sim 0$'' cross (as in Fig.~3)
	regions at very low gradient, where the values of $R_W-R_S$ is either
	zero or of the order of the numerical accuracy ($\lesssim 0.03$ SNU).}
The leading contribution to the winter-summer difference in the LOW  region
appears to be given by {\em pp\/} neutrinos, with a subleading contribution
from Be neutrinos, and negligible contributions from other sources. The $\delta
m^2$ value at the maximum of each plot roughly scales as $\delta m^2/\langle
E_\nu\rangle$, where $\langle E_\nu \rangle$ is the average neutrino energy;
therefore, the sum of the various contributions  produces separates peaks in
the total value of $R_W-R_S$, as already noted in Fig.~3.

Figure~4 demonstrates that the winter-summer asymmetry in GNO is mainly
sensitive to {\em pp\/} neutrinos, thus providing a test of the LOW solution
intrinsically different from the BOREXINO night-day asymmetry, which is
dominated by Be neutrinos. The two tests are also complementary in tracking the
origin of the time variations: a possible seasonal signal in GNO, if originated
by vacuum (instead of MSW) oscillations, should not produce a $N-D$ asymmetry
in BOREXINO.

It is curious to notice that the latitude dependence of the seasonal variations
is different from the case of day-night variations, which could be enhanced in
an equatorial experiment \cite{Ba97,Li97,Ge97}.  In fact, the difference
$R_W-R_S$ vanishes at the equator, while it becomes equal to the night-day rate
difference at the poles, where nighttime coincides with winter.

A final remark about uncertainties is in order. The ``theoretical'' error in
$R_W-R_S$ at GNO, due to systematic uncertainties of the SSM, is very small. In
fact, for a typical expected signal $R_W-R_S\simeq 4{\rm\ SNU\ ({\em pp})} + 1
{\rm \ SNU\ (Be)}$, the $1\sigma$ SSM errors \cite{BP98} amount to $\sim 0.04$
SNU and $\sim 0.09$ SNU for the {\em pp\/} and Be flux components,
respectively. Moreover, since the {\em pp\/} and Be flux errors are strongly
anticorrelated \cite{FoLi},  the two error components tend to cancel, giving a
total theoretical uncertainty as small as  $\sim 0.05$ SNU. Concerning the
experimental errors, they can only be guessed at present. The statistical
uncertainty of $R_W-R_S$, given the prospective estimates of \cite{GNOE},
should be $\lesssim 2$ SNU after about one solar cycle. The systematics might
be dangerously larger \cite{GNOE}; however, their constant components are
expected to cancel, to a large extent, in a rate difference such as $R_W-R_S$.
A dedicated  Monte~Carlo simulation would be very helpful to estimate the total
errors, and thus the statistical significance, of the $R_W-R_S$ measurement at
GNO.

\section{Summary and Prospects }

We have shown that the so-called ``LOW'' MSW solution to the solar neutrino
problem can be tested through Earth regeneration effects in two complementary
ways: day-night variations in BOREXINO and seasonal variations in GNO. The
first are dominated by Be neutrinos, and the second by {\em pp\/} neutrinos.
The net winter-summer difference  of event rates in GNO (removing eccentricity
effects) amounts to about 4--6 SNU within the LOW solution, and it might be
detected in GNO after $\sim 10$ years of data taking,  assuming the progressive
error reduction estimated in \cite{GNOE}.

In this work we have focussed on six-months averaged GNO rates, in order to
compute  the seasonal effect corresponding to the largest statistics. Of
course, the effect could be larger in selected (shorter) intervals of time, but
the corresponding statistics would also be lower. It might be useful to
investigate optimal periods for time averaging (maximizing the signal-to-error
ratio), when the prospective GNO statistical and systematic uncertainties will
be quantified more precisely.

\acknowledgments

One of us (D.M.) thanks Z.\ Berezhiani and the organizers of the Gran Sasso
Summer Institute ``Massive Neutrino in Physics and Astrophysics,'' where part
of this work was done, for kind hospitality. 

\newpage

\appendix
\section{Solar exposure during winter and summer}

In this Section we discuss our calculation of the average survival probability
$P_{\rm SE}(E_\nu)$ for a $\nu_e$ crossing the Sun (S) and the Earth (E),
during winter and summer time. The method is similar to the one used for yearly
averaging in Ref.~\cite{Li97}, to which we refer the reader for notation,
conventions, and details not repeated here.

We define the daily time $\tau_d$ as $2\pi\times {\rm day}/365$, so that
$\tau_d\in [0,2\pi]$, with $\tau_d=0$ at winter solstice (21 december).
``Winter'' (W) is then defined as a period of $\pm 3$ months $(\pm \pi/2)$
centered at $\tau_d=0$, and ``Summer'' (S) as the complementary period of the
year (see also Sec.~III). Since the physics is symmetric in the winter
half-intervals $[-\pi/2,0]$ and $[0,\pi/2]$, the time integration interval can
be reduced to $[0,\pi/2]$ for winter.

An important quantity, which depends on the latitude $\lambda$, is represented
by the fraction $\alpha$ of nighttime over the whole day during winter:
\begin{equation}
\alpha=\frac{2}{\pi^2}\int_0^{\pi/2}d\tau_d \,
\arccos[\tan\lambda\tan\delta_S(\tau_d)]\ ,
\end{equation}
where $\delta_{\rm S}$  is the sun declination,
\begin{equation}
\sin\delta_{\rm S}=-\sin i\,\cos\tau_d\ ,
\end{equation}
$i$ being the Earth inclination (0.4901 rad). At the Gran Sasso site 
($\lambda=42.45^\circ$) one gets $\alpha=0.5795$. The fraction of nighttime
during summer is, of course, $1-\alpha=0.4205$.

In order to calculate the average Sun-Earth survival probability $\langle
P_{\rm SE}\rangle_W$ over the winter period, it is useful to split such period
into nighttime (N) and daytime (D). During daytime, the Earth correction is
absent, and only the survival probability in the Sun ($P_{\rm S}$) is 
relevant:
\begin{eqnarray}
\langle P_{\rm SE} \rangle_W 
&=& \alpha \langle P_{\rm SE} \rangle_{W,N} +
(1-\alpha) \langle P_{\rm SE} \rangle_{W,D}\nonumber\\
&=& \alpha \langle P_{\rm SE} \rangle_{W,N}
+ (1-\alpha)\,P_{\rm S}\ .
\end{eqnarray}
Similarly, for summer averages one has:
\begin{equation}
\langle P_{\rm SE} \rangle_S = (1-\alpha )\langle P_{\rm SE} \rangle_{S,N}
+ \alpha\,P_{\rm S}\ ,
\end{equation}
so that the task is reduced to the calculation of the  winter and summer
averages during nighttime, $\langle P_{\rm SE}\rangle_{W,N}$ and $\langle
P_{\rm SE}\rangle_{S,N}$.

Following the approach of Ref.~\cite{Li97}, we transform the integrations over
winter  and summer nighttime in integrations over the nadir angle $\eta\in
[0,\pi/2]$ through appropriate weight functions $W_W(\eta)$ and $W_S(\eta)$:
\begin{eqnarray}
\langle P_{\rm SE}\rangle_{W,N} 
&=&\frac{\displaystyle\int^{\pi/2}_0 d\eta\,W_W(\eta)\,P_{\rm SE}(\eta)}
{\displaystyle\int^{\pi/2}_0\,d\eta\,W_W(\eta)}\ ,\\
\langle P_{\rm SE}\rangle_{S,N} 
&=&\frac{\displaystyle\int^{\pi/2}_0 d\eta\,W_S(\eta)\,P_{\rm SE}(\eta)}
{\displaystyle\int^{\pi/2}_0\,d\eta\,W_S(\eta)}\ .
\end{eqnarray}
The weight functions $W_W(\eta)$ and $W_S(\eta)$ simply represent the solar
exposure as a function of the nadir angle, namely, they are proportional to the
total fraction of time spent by the Sun as it repeatedly crosses  (from a
``Ptolemaic'' viewpoint)  a given nadir angle during winter and summer
nighttime, respectively.

In order to show the net seasonal effect due to MSW regeneration in the Earth,
we remove eccentricity effects by calculating the weight functions for a
``circular'' Earth orbit. Of course, this implies that the  eccentricity
($1/L^2$) effects shall be appropriately removed also from the experimental
rates. Notice that the data must be corrected in any case for time-variation
studies, in order to account for blank runs, different durations of runs, and
for runs crossing the winter-summer separation days (equinoxes). For a circular
earth orbit, it turns out that the weight  functions can be generally expressed
\cite{Li97} in terms of complete  and incomplete elliptic integrals of the
first kind,  $K$ and $F$ \cite{Grad}:%
\footnote{
	Alternatively, one could absorb the correction due to the eccentricity
	$\epsilon$ in the weight functions, and compare with the raw data. In
	this approach \cite{Ba97,500d}  (not adopted in this work), there would
	be a residual (geometrical) winter-summer asymmetry in the absence of
	neutrino oscillations, and the weight functions $W(\eta)$ should be
	corrected by an additional term $\pm \epsilon Y(\eta)$ containing
	elliptic integrals of the third kind; see Appendix~D of  
	\protect\cite{Li97}.
}
%
\begin{eqnarray}
K(x) &=& F(\pi/2,x)\, \\
F(\phi,x)&=&\int^{\sin\phi}_0 \frac{ds}{\sqrt{(1-s^2)(1-x^2\,s^2)}}\ .
\end{eqnarray}
Such integrals can be calculated numerically, e.g., through the routines  in
\cite{CERN}.

We give only the final results for the winter and summer weight functions
$W_W(\eta)$ and $W_S(\eta)$ at the Gran Sasso latitude, the derivation being
analogous to the case of the annual weight function $W(\eta)$ in \cite{Li97}:
\begin{equation}
W_{W}(\eta) = \frac{2\sin{\eta}}{\pi^2}\cdot\left\{ 
\begin{array}{ll}
0, 			& 0\leq \eta < \lambda-i, \\
z^{-1}\; K(y/z),	& \lambda-i \leq\eta \leq\lambda, \\
z^{-1}\,F(\chi_1,y/z),	& \lambda \leq \eta \leq \lambda+i,\\
y^{-1}\,F(\chi_2,z/y), 	& \lambda+i\leq\eta\leq\pi/2,
\end{array}
\right.\ 
\label{ww}
\end{equation}
\begin{equation}
W_{S}(\eta) = \frac{2\sin{\eta}}{\pi^2}\cdot\left\{ 
\begin{array}{ll}
0, 			& 0\leq \eta < \lambda, \\
z^{-1}\,F(\xi_1,y/z),	& \lambda \leq \eta < \lambda+i,\\
y^{-1}\,F(\xi_2,z/y), 	& \lambda+i < \eta\leq\pi/2\ ,
\end{array}
\right.\
\label{ws}
\end{equation}

In the above equations, the  arguments of the functions are defined as follows:
\begin{equation}
z = \sqrt{\sin i\,\cos\lambda\,\sin\eta}\ ,
\end{equation}
\begin{equation}
y = \sqrt{
\sin\left(\frac{i+\lambda+\eta}{2} \right)\,
\sin\left(\frac{i-\lambda+\eta}{2} \right)\,
\cos\left(\frac{i+\lambda-\eta}{2} \right)\,
\cos\left(\frac{i-\lambda-\eta}{2} \right)\,
}\ ,
\end{equation}
\begin{equation}
p = \sin(\lambda-\eta)/\sin i\ ,
\end{equation}
\begin{equation}
q = \sin(\lambda+\eta)/\sin i\ ,
\end{equation}
\begin{equation}
\chi_1 = \arcsin \sqrt{\frac{q-p}{(1-p)q}}\ ,
\end{equation}
\begin{equation}
\chi_2 = \arcsin \sqrt{\frac{1+q}{2q}}\ ,
\end{equation}
\begin{equation}
\xi_1 = \arcsin \sqrt{\frac{2p}{p-1}}\ ,
\end{equation}
\begin{equation}
\xi_2 = \arcsin \sqrt{\frac{p-1}{2p}}\ .
\end{equation}

The weight functions $W_W(\eta)$ and $W_S(\eta)$, defined in 
Eqs.~(\ref{ww},~\ref{ws}), have the following normalization:
\begin{eqnarray}
\int^{\pi/2}_0 d\eta\, W_W(\eta) &=& \alpha \,\\
\int^{\pi/2}_0 d\eta\, W_S(\eta) &=& 1-\alpha\ ,
\end{eqnarray}
and their sum, $W_Y(\eta)=W_W(\eta)+W_S(\eta)$, represents the weight function
appropriate for integration over the year.

Fig.~5 shows the winter, summer, and year weight functions  in terms of the
nadir angle $\eta$. Notice that the Earth core is crossed only during winter
time, and that trajectories are generally weighted in different ways during
winter and summer. This difference implies that the winter-summer asymmetry
does not simply depend on different fractions of nighttime in winter $(\alpha)$
and summer $(1-\alpha)$, as  derived in \cite{Sm99} in first approximation, and
explains why, in general, the day-night and winter-summer asymmetries are not
strictly proportional to each other \cite{Seas}.  Notice that the function
$W_Y(\eta)$ in Fig.~5 is identical to the function named $W(\eta)$ in
\cite{Li97} (for the Gran Sasso site).

Finally, the calculated values of $\langle P_{\rm SE} \rangle_{W,S}$ (for each
given neutrino energy $E_\nu$)  are used to compute the  winter and summer
event rates $R_{W,S}$ in GNO through the equations:
\begin{eqnarray}
R_W &=& \int dE_\nu \,\varphi(E_\nu)\, \sigma(E_\nu)\, 
\langle P_{\rm SE}(E_\nu)\rangle_{W}\ ,\\
R_S &=& \int dE_\nu \,\varphi(E_\nu)\, \sigma(E_\nu)\, 
\langle P_{\rm SE}(E_\nu)\rangle_{S}\ ,  
\end{eqnarray}
where $\varphi(E_\nu)$ is the solar neutrino energy spectrum \cite{BP98} and
$\sigma(E_\nu)$ is the neutrino absorption cross section in gallium
\cite{Gacs}.


\begin{figure}
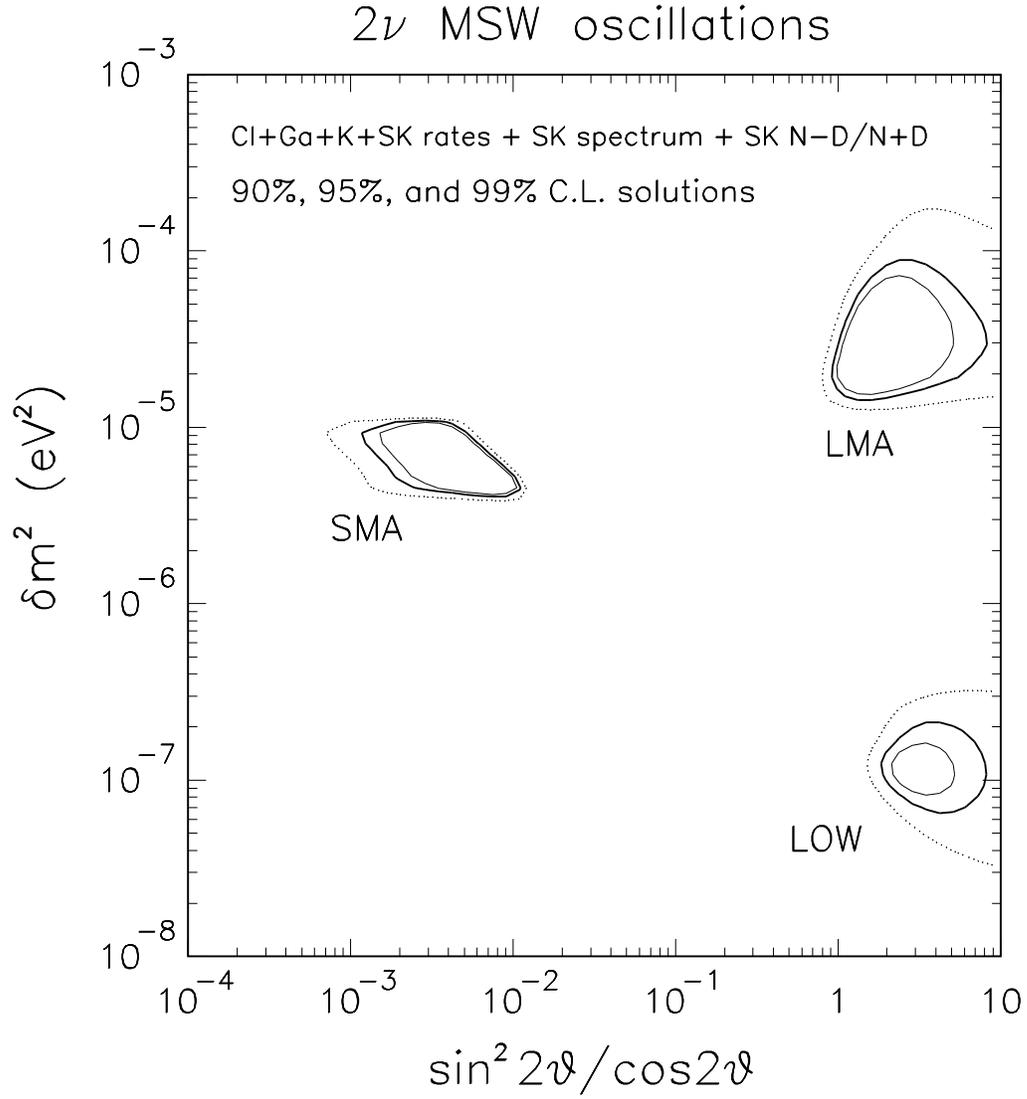

\caption{	Results of our global $\chi^2$  fit \protect\cite{Last}  for
		two-neutrino MSW oscillations, including total neutrino event
		rates from the Chlorine \protect\cite{Cl98}, Gallium
		\protect\cite{Ha99,Ab99},  and Water-Cherenkov experiments
		\protect\cite{Fu96,To99},  as well as the Super-Kamiokande
		night-day asymmetry and electron recoil energy spectrum
		\protect\cite{Su99}. Neutrino fluxes, spectra and cross
		sections are taken from  \protect\cite{BP98}.  The contours  at
		90\%, 95\%, and 99\% C.L.\ correspond to $\Delta \chi^2 =
		4.61$, 5.99, and 9.21.  
}
\label{fig1}
\end{figure}
\begin{figure}
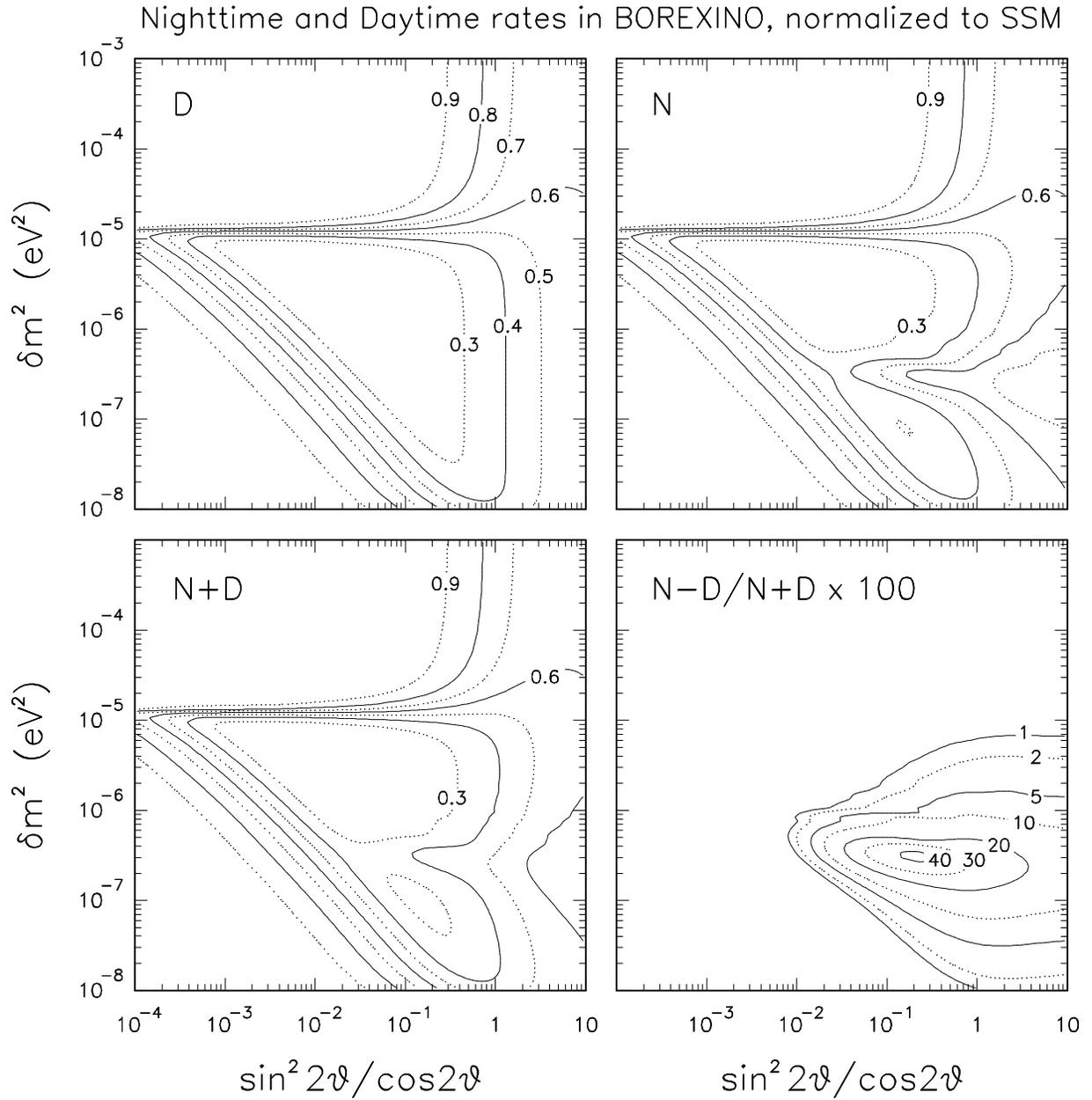

\caption{	Expected neutrino event rates in BOREXINO (averaged over one
		year) during daytime ($D$), nighttime ($N$), and full day
		($N+D$), together with the night-day asymmetry $N-D/N+D$. Rates
		are normalized to unoscillated expectations.
}
\label{fig2}
\end{figure}
\begin{figure}
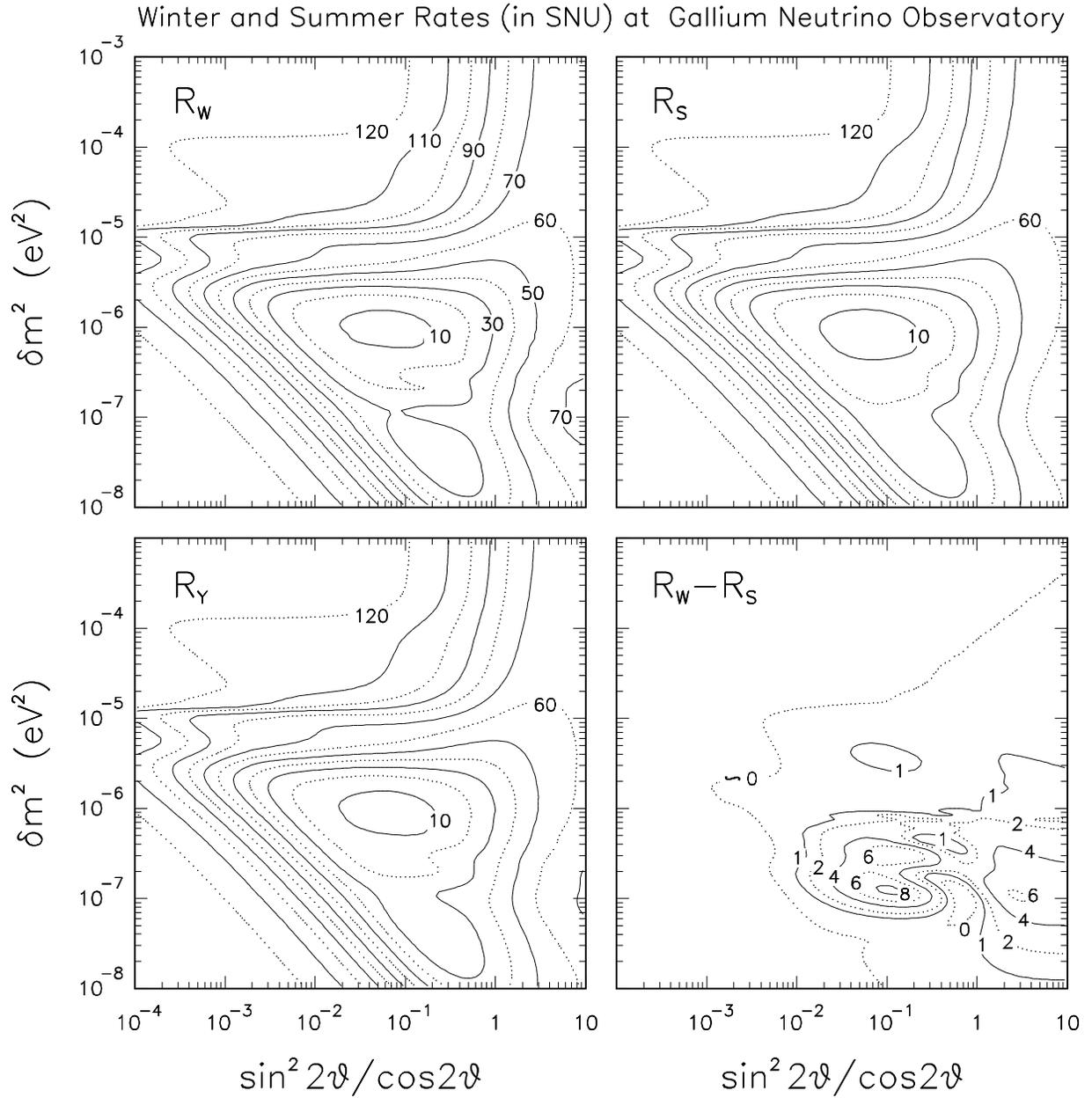

\caption{	Expected neutrino event rates $R$ in GNO, averaged over 
		``Winter''  [23 sep.--21 mar.], ``Summer'' [22 mar.--22 sep.],
		and Year, together with the winter-summer difference $R_W-R_S$.
		All rates are given in SNU. Eccentricity effects are removed,
		so that $R_W-R_S$ represents the net MSW seasonal effect due
		$\nu_e$ regeneration in the Earth.
}
\label{fig3}
\end{figure}
\begin{figure}
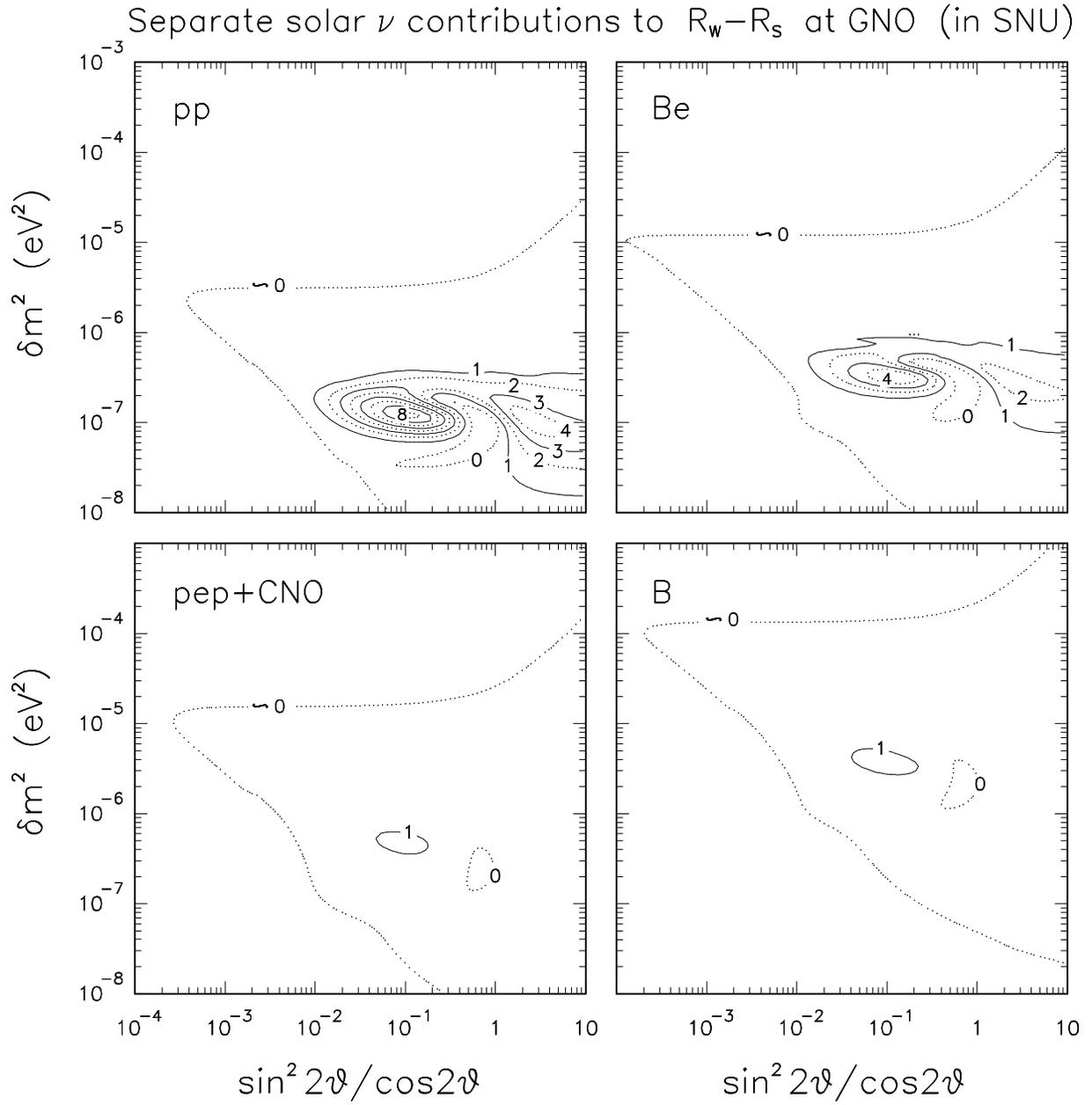

\caption{	Separate contributions to the winter-summer difference 
		$R_W-R_S$ in GNO due to {\em pp}, {\em pep}+CNO, and B solar
		neutrinos. Isolines correspond to integer values of 
		$(R_W-R_S)/$SNU in steps of 1~SNU (dotted lines: even values;
		solid lines: odd values).
}
\label{fig4}
\end{figure}
\begin{figure}
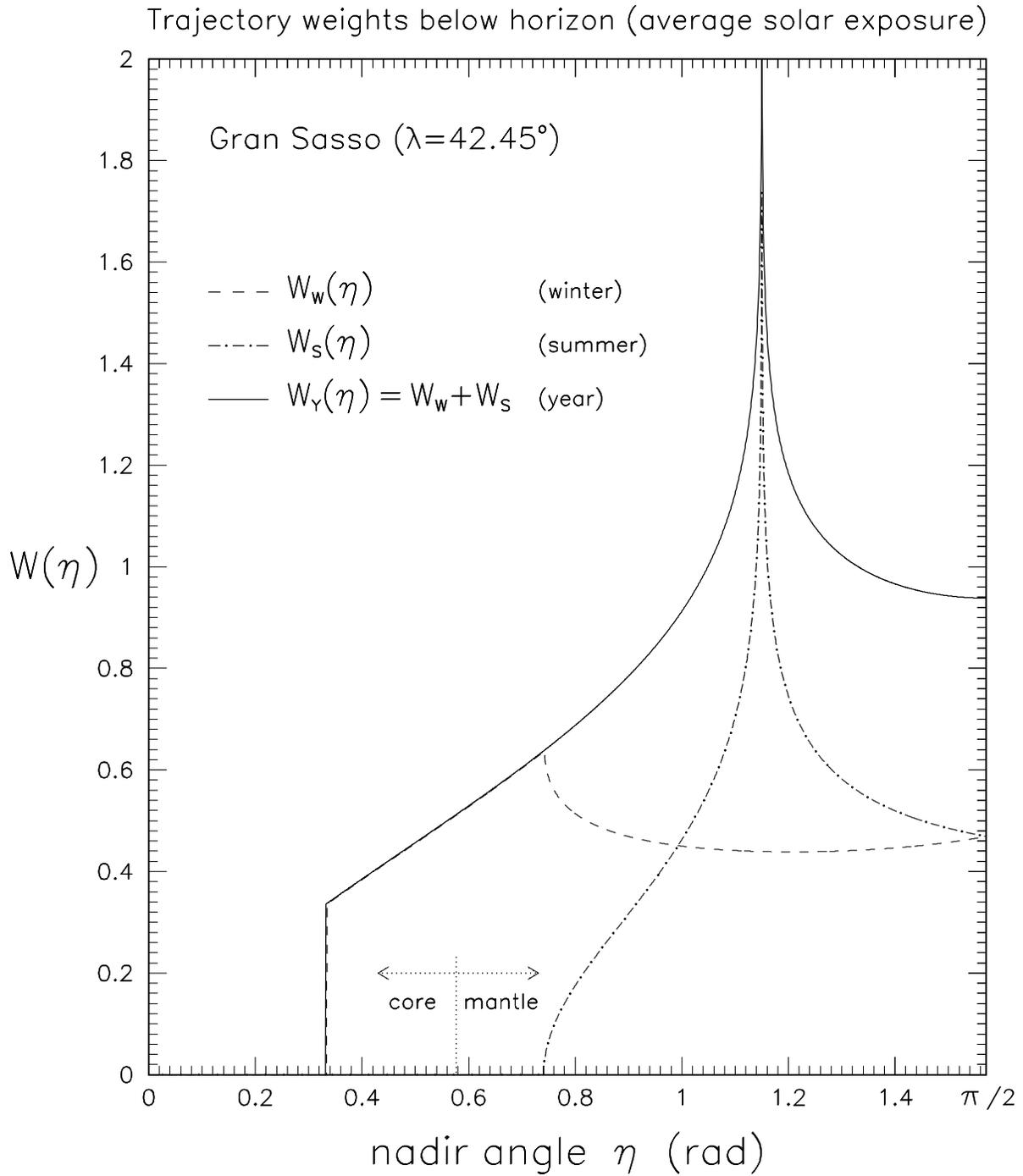

\caption{	Solar exposure functions {\em vs} nadir angle during winter,
		summer, and the whole year (at the Gran Sasso site). The
		functions $W_Y$ and $W_W$ coincide for $\eta <\lambda$.
}
\label{fig5}
\end{figure}


\newcommand{\InsertFigure}[2]{\newpage\phantom{.}
\vspace*{-2.cm}\begin{center}\mbox{%
\epsfig{bbllx=2truecm,bblly=2truecm,bburx=19.5truecm,bbury=26.5truecm,%
height=23.truecm,figure=#1}}\end{center}\vspace*{-2.85truecm}%
\parbox[t]{\hsize}{\small\baselineskip=0.5truecm\hskip0.5truecm #2}}

\InsertFigure{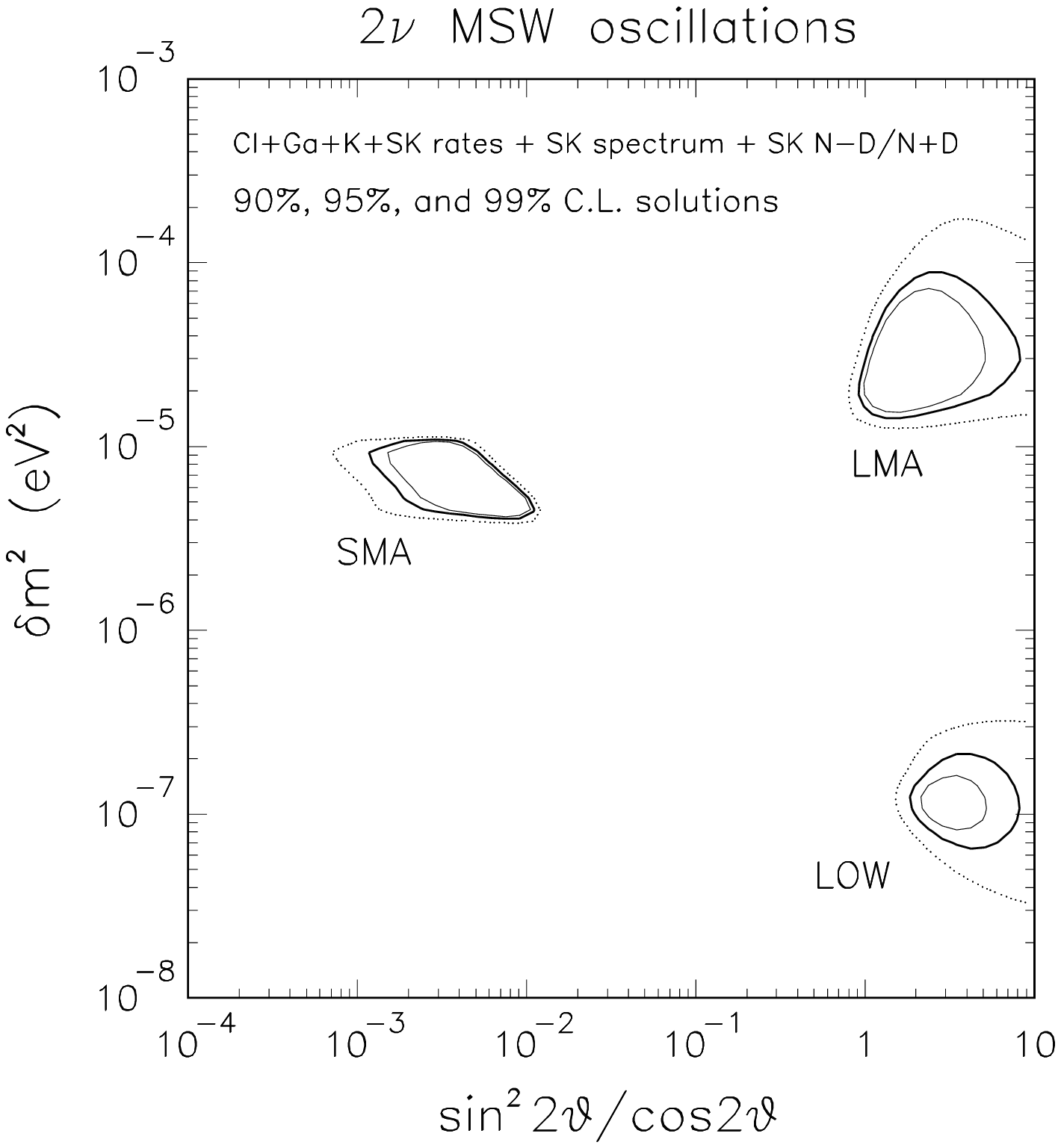}%
{FIG.~1.	Results of our global $\chi^2$  fit \protect\cite{Last}  for
		two-neutrino MSW oscillations, including total neutrino event
		rates from the Chlorine \protect\cite{Cl98}, Gallium
		\protect\cite{Ha99,Ab99},  and Water-Cherenkov experiments
		\protect\cite{Fu96,To99},  as well as the Super-Kamiokande
		night-day asymmetry and electron recoil energy spectrum
		\protect\cite{Su99}.  Neutrino fluxes, spectra and cross
		sections are taken from  \protect\cite{BP98}.   The contours at
		90\%, 95\%, and 99\% C.L.\ correspond to $\Delta \chi^2 =
		4.61$, 5.99, and 9.21. 
}
\InsertFigure{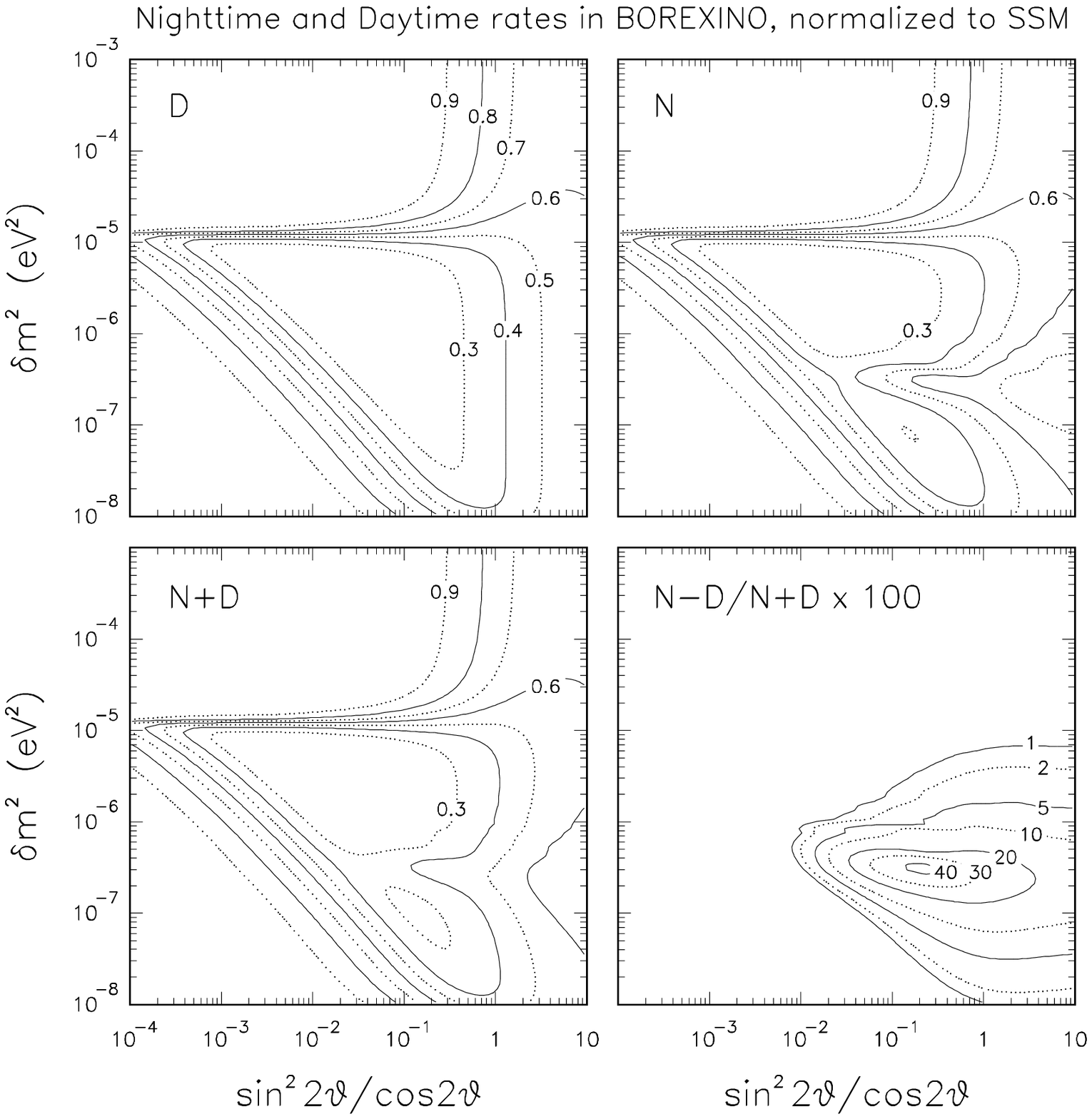}%
{FIG.~2.	Expected neutrino event rates in BOREXINO (averaged over one
		year) during daytime ($D$), nighttime ($N$), and full day
		($N+D$), together with the night-day asymmetry $N-D/N+D$. Rates
		are normalized to unoscillated expectations.	
}
\InsertFigure{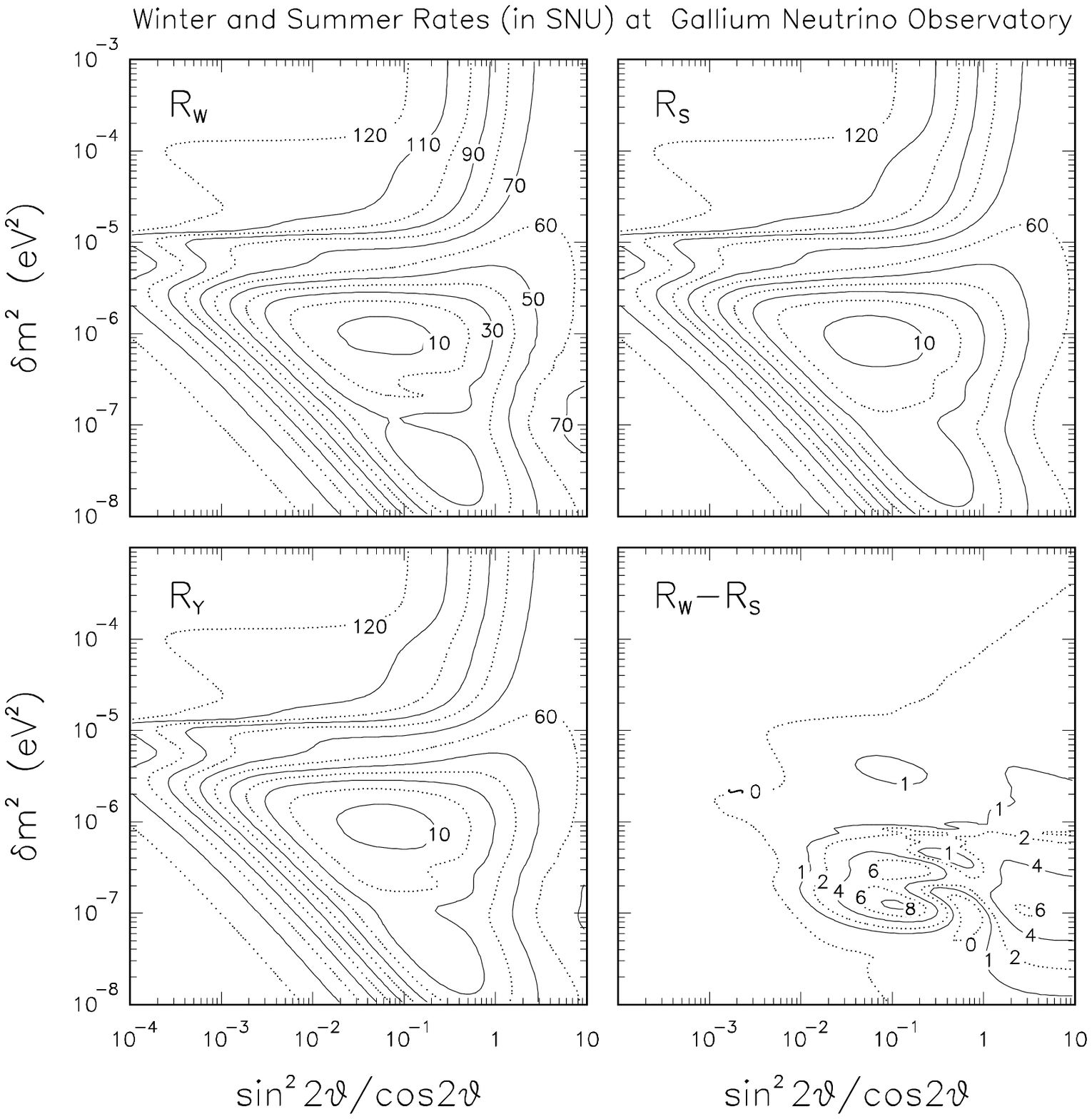}%
{FIG.~3.	Expected neutrino event rates $R$ in GNO, averaged over 
		``Winter'' [23 sep.--21 mar.], ``Summer'' [22 mar.--22 sep.],
		and Year, together with the winter-summer difference $R_W-R_S$.
		All rates are given in SNU. Eccentricity effects are removed,
		so that $R_W-R_S$ represents the net MSW seasonal effect due to
		$\nu_e$ regeneration in the Earth.
}
\InsertFigure{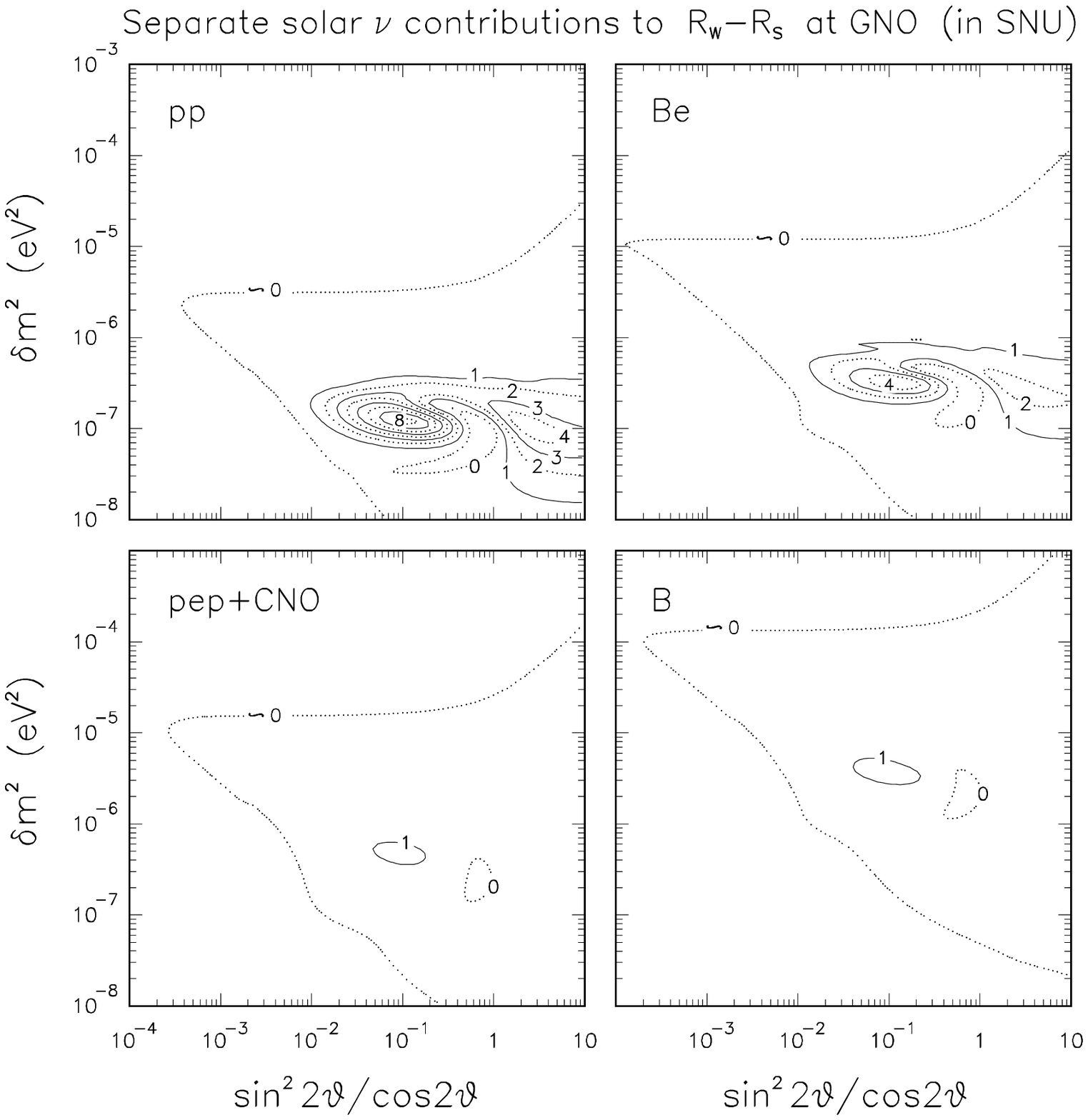}%
{FIG.~4.	Separate contributions to the winter-summer difference 
		$R_W-R_S$ in GNO due to {\em pp}, {\em pep}+CNO, and B solar
		neutrinos. Isolines correspond to integer values of 
		$(R_W-R_S)/$SNU in steps of 1~SNU (dotted lines: even values;
		solid lines: odd values).
}
\InsertFigure{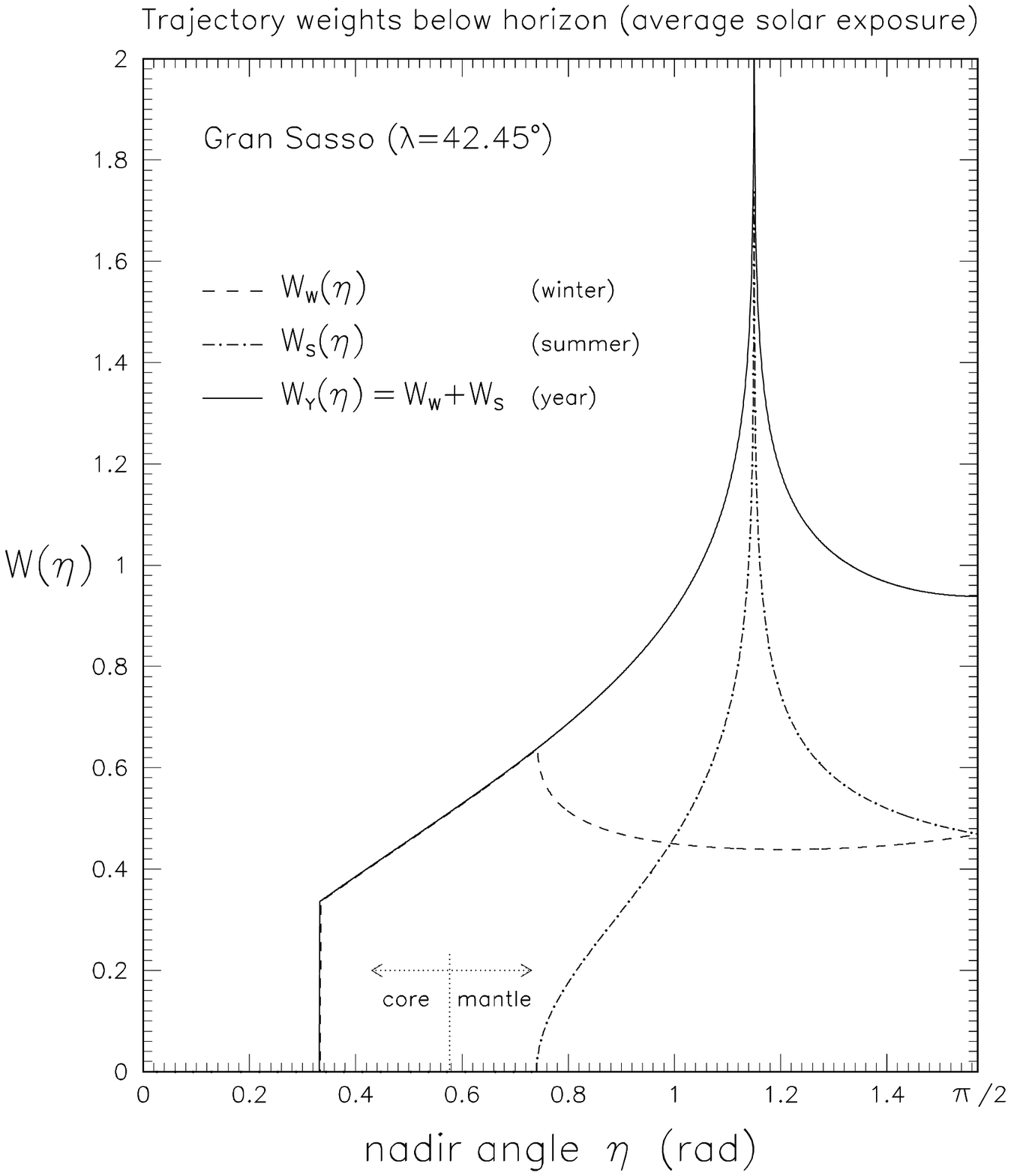}%
{FIG.~5.	Solar exposure functions {\em vs} nadir angle during winter,
		summer, and the whole year (at the Gran Sasso site). The
		functions $W_Y$ and $W_W$ coincide for $\eta <\lambda$.
}


\end{document}